\documentclass[pre,aps,superscriptaddress,amsmath,amssymb,showpacs,twocolumn,floatfix]{revtex4}
\usepackage{hyperref}
\usepackage{graphicx}
\usepackage{amsmath}
\usepackage{amssymb}

\newcommand{\rem}[1]{}

\begin{document}

\title{Influence of substrate potential shape on the dynamics of a sliding lubricant chain}
\author{Rosalie Laure Woulach\'{e}}
\affiliation{
Laboratoire de M\'{e}canique, D\'{e}partement de
Physique, Facult\'e des Sciences,
Universit\'e de Yaound\'e I. B.P. 812,
Yaound\'{e}, Cameroun}
\affiliation{The Abdus Salam International Centre for Theoretical Physics,
Strada Costiera, 11-34014 Trieste, Italy}
\email{rwoulach@yahoo.com; rwoulach@ictp.it}

\author{Andrea Vanossi}
\affiliation{CNR-IOM Democritos National Simulation Center, Via Bonomea
  265, 34136 Trieste, Italy}
\affiliation{International School for Advanced Studies (SISSA), Via Bonomea
 265, 34136 Trieste, Italy}

\author{Nicola Manini}
\affiliation{CNR-IOM Democritos National Simulation Center, Via Bonomea
  265, 34136 Trieste, Italy}
\affiliation{International School for Advanced Studies (SISSA), Via Bonomea
 265, 34136 Trieste, Italy}
\affiliation{Dipartimento di Fisica, Universit\`a degli Studi di
  Milano, Via Celoria 16, 20133 Milano, Italy}

\date{\today}

\begin{abstract}
We investigate the frictional sliding of an incommensurate chain of
interacting particles confined in between two nonlinear on-site substrate
potential profiles in relative motion.
We focus here on the class of Remoissenet-Peyrard parametrized potentials
$V_{\rm RP}(x,s)$, whose shape can be varied continuously as a function of
$s$, recovering the sine-Gordon potential as particular case.
The observed frictional dynamics of the system, crucially dependent on the
mutual ratios of the three periodicities in the sandwich geometry, turns
out to be significantly influenced also by the shape of the substrate
potential.
Specifically, variations of the shape parameter $s$ affects significantly
and not trivially the existence and robustness of the recently reported
velocity quantization phenomena [Vanossi {\it et al.},
  Phys.\ Rev.\ Lett.\ {\bf 97}, 056101 (2006)], where the chain
center-of-mass velocity to the externally imposed relative velocity of the
sliders stays pinned to exact ``plateau'' values for wide ranges of the
dynamical parameters.
\end{abstract}


\maketitle

\section{Introduction}

Sliding friction has been a broadly-studied field due to its huge
practical relevance as well as its theoretical challenges
\cite{Bowden50,PerssonBook}.
The regime of validity and the microscopic origin of the Amontons-Coulomb
empirical laws of static and dynamic friction have still open issues
\cite{Rubinstein04,Rubinstein06}.
The advancements of technology in the last few decades has triggered both
theoretical
\cite{Robbins01,Braun06,molshapeLin,Bonelli09,Negri10,Braun13,VanossiRMP13}
as well as experimental \cite{Carpick97,Dienwiebel04,Krim12} investigations
in this field.

A broad range of investigations focuses on a simple fundamental model for
microscopic tribological system: the Frenkel-Kontorova (FK) model
\cite{Floria96} and its extensions \cite{Braunbook}.
Its standard 1D version consists of a chain of harmonically interacting
atoms subject to one periodic sinusoidal potential, thereby representing
a discretized elastic overlayer deposited on a corrugated surface.
The application of a constant force to the chain allows us to determine a
depinning threshold, representative of static friction.
For an irrational ratio between the natural atomic spacing and the period
of the substrate potential (incommensurate interface), the FK model
undergoes a phase transition, called Aubry transition, where the ground
state ``hull function'' exhibits analyticity breaking \cite{Peyrard83}.
When, for a fixed interparticle chain stiffness, the amplitude of the
sinusoidal potential is smaller than a certain critical value, the static
frictional force vanishes, leading to the onset of free sliding or
``superlubricity'';
otherwise the chain is pinned until a finite threshold force is overcome.

Superlubricity connected with incommensurability is one of the pervasive
concepts of modern tribology with a wide area of relevant practical
applications as well as fundamental theoretical issues
\cite{Erdemir07}.
%
The role of incommensurability has been recently extended \cite{Braun05} in
the framework of a driven 1D confined model inspired by the tribological
problem of two sliding interfaces with a thin lubricant layer in between.
The moving interface is thus characterized by three inherent length scales:
the periods of the bottom and top substrates, and the period of the
embedded solid lubricant structure.
In particular, in the presence of a uniform external driving of the top
substrate, the interplay between these incommensurate length scales can
give rise to intriguing dynamical phase-locking phenomena and surprising
velocity ``quantization'' effects, due to the dragging of topological
solitons (``kinks'' and ``antikinks''), i.e., nonlinear localized density
superstructures arising from geometrical lattice mismatch \cite{Vanossi06,
  Santoro06, Cesaratto07, Vanossi07Hyst, Manini07extended,
  Vanossi08TribInt}.

These results are suggestive but remain rather idealized in several
respects.
In particular, the profile of the corrugation potential energy experienced
by a lubricant atom interacting with real physical surfaces is likely to
deviate considerably from the sinusoids of the two-substrate confined
tribological model.
It is therefore useful to investigate what influence the shape of the
substrate corrugation may have on the frictional dynamics.
In this paper we model the corrugation of the two confining substrates via
the Remoissenet-Peyrard (RP) function \cite{Remoissenet81,Peyrard82}, whose
shape can be varied continuously as a function of a parameter.
The RP potential, which retrieves the sine-Gordon shape as a special case,
has been employed widely and successfully to model the dynamics of atoms
adsorbed on crystal surfaces in realistic situations
\cite{Braunbook,DjuidjeKenmoe04,HuTekic05,Woulache05}.

\section{The model}

\begin{figure}
  \begin{center}
  \includegraphics[width=80mm,clip=]{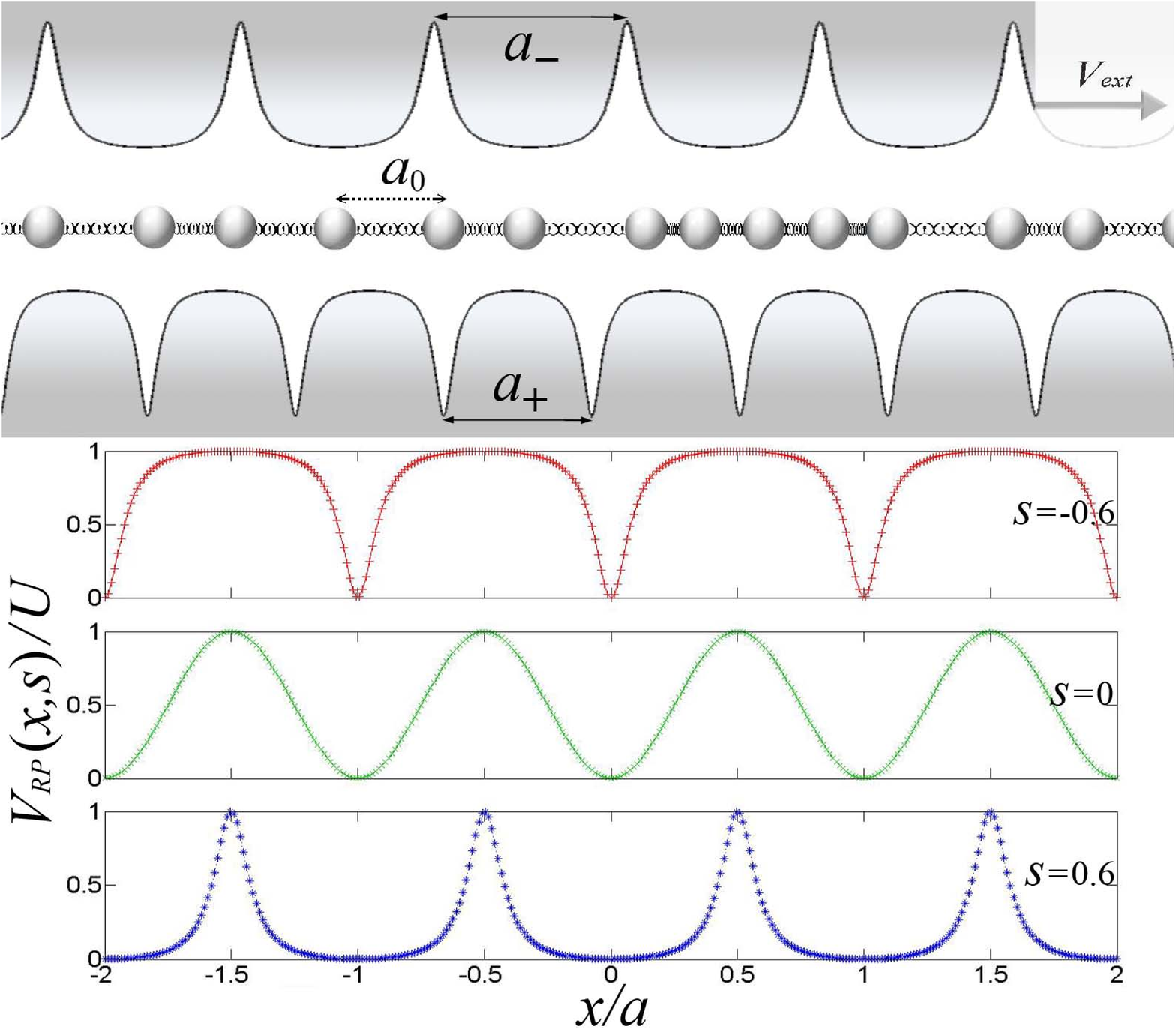}
  \end{center}
\caption{\label{model:fig} (Color online)
Scheme of the model with three characteristic length scales $a_+$ (spatial
period of the static substrate), $a_0$ (the equilibrium spacing of the
harmonic chain representing the lubricant film), and $a_-$ (spatial period
of the advancing substrate).
The substrate corrugation is modeled by the RP potential, which is
illustrated below for three values of the deformation parameter.
Top to bottom $s=-0.6$ (narrow valleys), $0$ (pure sine), and $0.6$
(broad valleys).
}
 \end{figure}

Consider the one-dimensional generalization of the two-sines FK model, as
in Ref.~\cite{Vanossi06}, consisting here of two RP substrate potentials,
of spatial periodicity $a_{+}$ and $a_{-}$ and a chain of interacting
particles of mass $m$ and harmonic spring constant $K$, equally spaced by a
lattice constant $a_{0}$, mimicking the sandwiched lubricant layer as shown
in Fig.~\ref{model:fig}.
The motion of the $i$-th lubricant particle is governed by
\begin{eqnarray}\label{model:eq}
 m\ddot{x_{i}} &=& - \frac{dV_{\rm RP+}(x_i-v_+ t,s)}{dx}
                   - \frac{dV_{\rm RP-}(x_i-v_- t,s)}{dx}
\\\nonumber
&&+  K\left(x_{i+1}+x_{i-1}-2x_i\right)
 - \gamma\left(2 \dot x_i-v_+-v_-\right)
\,.
\end{eqnarray}
Here the RP potential \cite{Remoissenet81} $V_{\rm RP}(x,s)$ is defined by
\begin{equation}\label{eq2}
V_{\rm RP\pm}(x,s)  =\frac{U}{2}\,
\frac{(1-s)^{2}\left[1-\cos \left(k_{\pm}x\right)\right]}
{1+s^{2}+2s\cos\left(k_{\pm}x\right)}
\,,\quad |s|<1.
\end{equation}
For $s = 0$, the potential $V_{\rm RP\pm}(x,s)$ of amplitude $U$ yields a
sinusoidal shape; for $s>0$, it provides an array of broad wells separated
by narrow barriers; for $s<0$, it provides deep narrow wells separated
by broad flat barriers (see Fig.~\ref{model:fig}).
$k_{\pm} = 2\pi/a_{\pm}$ are the wave-vector periodicities associated to
the upper ($-$) and lower ($+$) substrates, and $v_{-}$, $v_{+}$ denote
their sliding velocities respectively.
$\gamma$ is a viscous-friction damping which takes into account various
sources of dissipation in the substrates (phonons, electronic excitations,
etc.), which are not explicitly included in the model.
We select a relatively small dissipation constant $\gamma=0.1$, producing
an underdamped regime.
As done in previous work \cite{Braun05, Vanossi06} to simulate an infinite
system, periodic boundary conditions (PBC) are applied, implementing,
e.g.\ via a continued fraction expansion technique \cite{Khinchin},
suitable rational approximations of the system periodicities $a_+$, $a_0$,
and $a_-$, mimicking the desired incommensurability.

As found in earlier studies, the general behavior of this model
depends crucially on the relative commensurability of the substrates and
the chain (see Ref.~\cite{Manini07PRE} and references therein).
To make a comparison with previous work in the sine-Gordon substrates, we
consider here the corresponding ratios of length scales defined by
$r=a_+/a_0 = N_0/N_+$ and $r'=a_-/a_+ = N_+/N_-$.
$r$ and $r'$ are also expressed in terms of the number of lubricant
particles $N_0$ and the numbers of periods $N_\mp$ of the top/bottom
potential oscillations in Eq.~\eqref{model:eq} within each simulation cell.
For definiteness we take $r'>r^{-1}$ and $r'>1$ so that the top substrate
has the longest lattice spacing.
We take $a_+ = 1$, $m = 1$, and the force $F_+=2\,\pi\, U/a_+$ as basic
units for the model.
In the following we express all physical quantities in terms of suitable
combinations of $a_+$, $m$, $F_+$.

An adaptive fourth-order Runge-Kutta algorithm is used to integrate the
equations of motion.
We start off with the chain particles at equilibrium (the local energy
minimum obtained by relaxing the immobile -- $v_\pm=0$ -- system from a
chain at rest with uniform separation $a_0$).
Without loss of generality, we select a reference frame such that the
bottom substrate is at rest ($v_+ = 0$), and make the upper substrate slide
at constant velocity $v_- = V_{\rm ext}$.
After an initial transient, the system reaches a dynamical steady state
characterized by regular or irregular fluctuations of the drift velocity of
the chain particles around an average value, which we indicate by $V_{\rm CM}$.

In the present work, we investigate the effects, on the tribological
behavior of the sliding interface, of ({\it i}) the shape of the substrates
potentials, represented by the RP parameter $s$, and ({\it ii}) the
coverage $\Theta$ of the upper substrate to the array of kinks or antikinks.
Since the mean distance between consecutive kink/antikinks is
\begin{equation}\label{dkink}
d_{\rm kink}= \frac 1{\frac 1{a_0} - \frac 1{a_+}} = \frac{a_0}{1 - r^{-1}}
\,,
\end{equation}
this coverage
\begin{equation}\label{coverage}
\Theta = \frac {a_-}{d_{\rm kink}} = r'\, \left|r - 1\right|
\end{equation}
can be tuned by modifying $r'$ \cite{Manini07extended, Manini07PRE}.

\section{Results and discussion}

\subsection{Kinks: lubricant forward motion}

\begin{figure}
  \begin{center}
   \includegraphics[width=80mm,clip=]{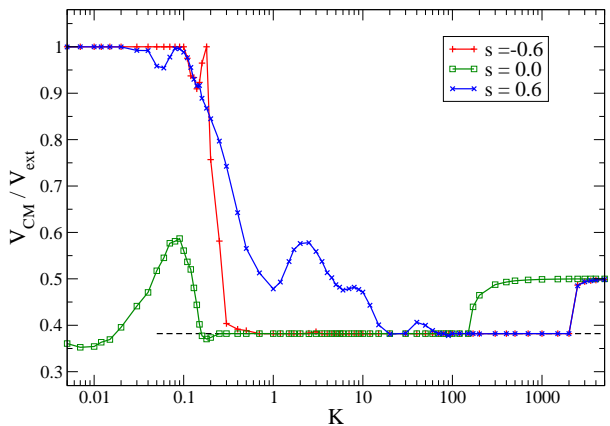}
  \end{center}
\caption{\label{GM1:fig} (Color online)
  Ratio $V_{\rm CM}/V_{\rm ext}$ of the steady-state average chain
  center-mass velocity to the top speed, as a function of the chain
  stiffness $K$, for three values $s = -0.6, 0.0, 0.6$ of the potential
  shape parameter.
  The length ratios are taken as approximants to the golden mean
  $r=377/233$, $r'=233/144$; the simulation involves $N_0=377$ particles; the
  damping and top driving speed are $\gamma=0.1$, $V_{\rm ext}=0.1$.
  The dashed line marks $w_{\rm plateau}\simeq 0.382$.
}
 \end{figure}

Figure~\ref{GM1:fig} reports the time-averaged center-of-mass velocity
$V_{\rm CM}$ of the chain (normalized to the top driving velocity $V_{\rm
  ext}$) as a function of its stiffness $K$, for three values of the
deformation parameter $s$.
In this calculation, for the length ratios we take a rational approximant
to the golden mean $r=\phi=(1+\sqrt{5})/2\simeq 1.62$.
We adopt $\Theta=1$, i.e.\ as many solitons as oscillation periods of the
top substrate, and this, due to Eq.~\eqref{coverage}, implies $r'=1/|r-1|$.
For $r\simeq \phi$, $r'$ happens to approximate $\phi$ as well.
According to Eq.~\eqref{coverage}, this choice for $r'$ produces a coverage
$\Theta=1$, i.e.\ as many solitons as oscillation periods of the top
substrate.
The velocity of the sliding top substrate is set to a moderate $V_{\rm ext}
= 0.1$.

Within a broad range of stiffness values, the chain moves at the quantized
velocity $V_{\rm CM} = w_{\rm plateau} V_{\rm ext}$ \cite{Vanossi06}, with
\begin{equation} \label{vplateau}
w_{\rm plateau} = 1-\frac 1r,
\end{equation}
approximately $w_{\rm plateau}\simeq 0.382$ for the adopted value of $r$.
For extremely soft harmonic interparticle couplings (small chain stiffness
$K$) and for $s = -0.6$ or $s = 0.6$, the chain center of mass tends to
move ahead at the full external velocity $V_{\rm ext}$.
In the opposite limit of a very stiff chain (large $K$), it moves at the
symmetric speed $V_{\rm ext}/2$.
This is expected in a situation where the chain-corrugation interaction
become marginal, and the dominating term in Eq.~\eqref{model:eq} is the
dissipative one, which is minimum when $\dot x_i = (v_++v_-)/2 =
V_{\rm ext}/2$.

In the transitions between the plateau speed and the large-$K$ and small-$K$
regimes, the chain average velocity $V_{\rm CM}$ is generally a nontrivial
function of the chain stiffness $K$.
The effect of the shape of the corrugation potential is evident: for
both positive and negative $s$, the plateau shrinks in size at the
soft-chain side (small $K$), while it tends to expand in the stiff-chain
side (large $K$).
While the large-$K$ expansion is the same for positive and negative $s$,
the small-$K$ shrinkage is far more dramatic for positive $s$, i.e.\ for
broad shallow minima separated by narrow sharp maxima in the corrugation
potential.
Such a behavior can be qualitatively understood by considering that the
plateau mechanism has been interpreted in terms of solitons, formed by the
mismatch of the chain periodicity to that of the more commensurate
substrate (here the bottom potential), being rigidly driven forward by the
(top) advancing potential representing the other, more mismatched,
sliding surface.
As $K$ is decreased, kinks become more and more localized objects: the
plateau ends when the Peierls-Nabarro barrier \cite{Floria96} for a kink to
move forward one lattice parameter approaches the single-particle activation
energy to jump a corrugation potential barrier.
When the two barriers coincide, no kink motion is granted any more, and
the chain advances as a whole ($V_{\rm CM}=V_{\rm ext}$).
For positive $s$ values, the possibility for the particles to arrange
relatively uniformly over the RP substrate at a small energy cost yields
poorly localized kink superstructures in the chain, rapidly becoming
equivalent to non-interacting particles, which are easily dragged forward,
and drag the whole chain along. This rapidly destroys the quantized
velocity plateau.
In contrast, the deep narrow wells for negative values of the shape
parameter $s$ tend to compress the chain particles in sharper kink
structures relatively more easily dragged along by the moving substrate,
while leaving the other (non-soliton) particles still pinned in the other
deep minima, and preserving the quantized motion down to softer $K$.
The potential deformation is beneficial in the rigid-chain regime because,
for given corrugation amplitude $U$, the maximum force (the slope at the
inflection point, see Fig.~\ref{model:fig}) that the top potential can
apply to the chain particles is larger for $s\neq0$ than for the harmonic
chain $s=0$.
The dragging force acting on the kinks is proportionally larger, allowing
dragging to extend to stiffer chains which come with broader and fainter
kinks.

\begin{figure}
 \begin{center}
  \includegraphics[width=80mm,clip=]{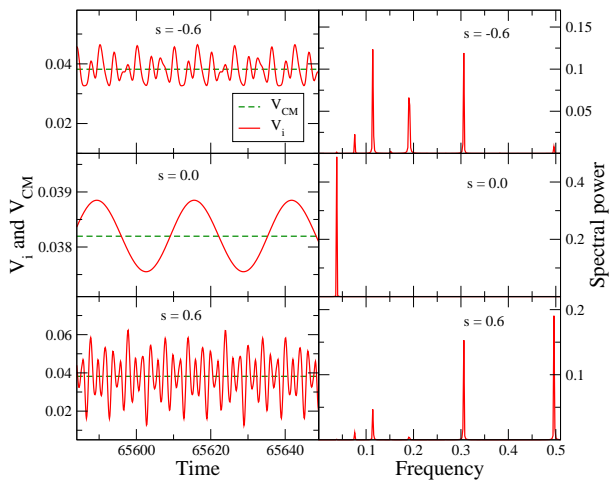}
\caption{\label{CMFT:fig} (Color online)
  Left panels: time-evolution of the velocity of a particle $V_i$ (solid)
  and the chain overall center of mass velocity $V_{\rm CM}$ (dashed), for
  three different shapes of the RP substrates potentials, and for stiffness
  $K=100$ ($s=0$) and $K=400$ ($s=\pm 0.6$), representative of the
  quantized plateau.
  Right panels: the corresponding Fourier analysis of the velocity signal,
  representing the power spectral decomposition of the motion.
}
 \end{center}
\end{figure}

To obtain a microscopic view of the quantized phenomena, we follow the
motion of a particle of the chain and plot its velocity $\dot x_i$, aside
with the CM velocity $V_{\rm CM}$, as a function of time (left panels of
Fig.~\ref{CMFT:fig}).
We compare the usual three values of the deformation parameter considered
in simulations.
We adopt relatively rigid stiffness $K = 100$ ($s=0$) or $400$ ($s=\pm0.6$)
selected to remain well inside the quantized plateau in all cases.
The motion of a single particle is a periodic oscillation, representing the
passage of a soliton across that specific particle.
This period $\tau$ equals the distance $d_{\rm kink}$ between successive
kinks divided by the speed $V_{\rm ext}$ at which they are dragged forward by
the advancing top substrate: $\tau = d_{\rm kink} V_{\rm ext}^{-1} = a_+
V_{\rm ext}^{-1} \left(1-1/r\right)^{-1}$.
%
%
Of course, this period is independent of the deformation parameter of the
substrates.
A Fourier analysis (right panels of Fig.~\ref{CMFT:fig}) reveals that
indeed the single-particle motion is periodic, with the same period
$\tau=1/\nu_0\simeq 26.1$, where $\nu_0\simeq 0.0382$ is the fundamental
harmonic peak frequency in the Fourier spectrum of the present examples.
However, the detailed motion induced by the deformed potential is clearly
very different, characterized by a remarkably high harmonic contents,
compared to the simple harmonic oscillation of the $s=0$ case.
The $s=\pm 0.6$ potential requires a complicated ``dance'' of the
individual atoms to accommodate the passage of a soliton.
Note also that the concerted oscillation of all particles in the chain
makes its center of mass advance at an essentially constant speed (dashed
line) $V_{\rm CM}(t) \equiv w_{\rm plateau}\,V_{\rm ext}$ within numerical
error.

\begin{figure}
 \begin{center}
  \includegraphics[width=80mm,clip=]{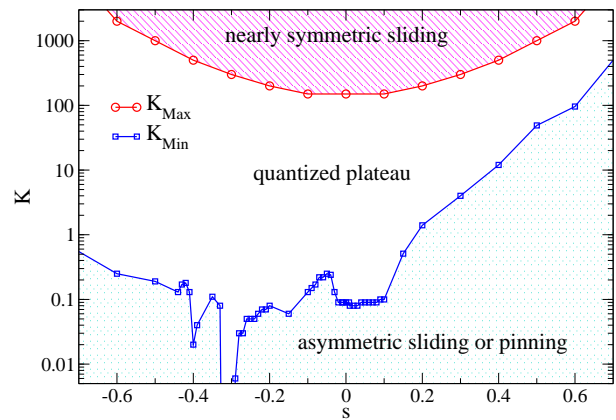}
\caption{\label{KmaxKmin:fig} (Color online)
 The boundaries $K_{\rm Min}$ and $K_{\rm Max}$ of the quantized velocity
 plateau as a function of the chain stiffness $K$, for varied potential
 shape parameter $s$.
 The model conditions are the same as in Fig.~\ref{GM1:fig}.
}
 \end{center}
\end{figure}

To characterize in greater detail the effect of the potential shape on the
quantized motion, we investigate the upper and lower boundary of the
quantized plateau, $K_{\rm Max}$ and $K_{\rm Min}$ respectively.
These boundaries are obtained by a sequence of linked calculations carried
out with increasing (for $K_{\rm Max}$) or decreasing (for $K_{\rm Min}$)
$K$ in small steps, until the quantized plateau is abandoned.
For example, the sequence of calculations of Fig.~\ref{GM1:fig} shows that
$K_{\rm Max}\simeq 2000$ for $s=\pm 0.6$.
Figure~\ref{KmaxKmin:fig} reports the dependency of $K_{\rm Min}$ and
$K_{\rm Max}$ on the potential shape parameter $s$.
$K_{\rm Max}$ is a symmetric function of the shape parameter $s$.
In contrast, the $K_{\rm Min}$ curve is quite asymmetric.
As already remarked above, at the soft-chain side positive $s$ is
consistently detrimental to the plateau state, leading to a rapid
(approximately exponential) increase of $K_{\rm Min}$ with $s$.
In contrast, the negative-$s$ region has a range $-0.4<s<-0.2$ where the
potential shape deformation is beneficial to the quantized plateau even for
soft chains.
A further decrease of $s$ to more negative values produces an increase of
$K_{\rm Min}$, but a slow one, such that the relative width $K_{\rm
  Max}/K_{\rm Min}$ of the plateau actually increases as $s$ decreases.
The $K_{\rm Min}$ and $K_{\rm Max}$ curves delimit the quantized velocity
plateau region in the space of parameters $s$ and $K$.
Above this region, we find a stiff-chain region where the dynamics is
dominated by the dissipative term in Eq.~\eqref{model:eq}, and $V_{\rm CM}$
approaches rapidly $V_{\rm ext}/2$, with the two sliders acting
symmetrically on the chain.
The soft-chain region below the $K_{\rm Min}$ exhibits occasional pinning
to either the top or the bottom slider, or unpinned nonquantized
nonperiodic orbits.

The ``dynamical phase diagram'' of Fig.~\ref{KmaxKmin:fig} is relevant for
the specific adopted value of dissipation $\gamma$ and of speed $V_{\rm
  ext}$.
A modification of these two parameters would indeed modify the shape of the
diagram, while preserving its overall features.

\subsection{Antikinks: lubricant backward motion}
\label{anti:sec}

\begin{figure}
  \begin{center}
   \includegraphics[width=80mm,clip=]{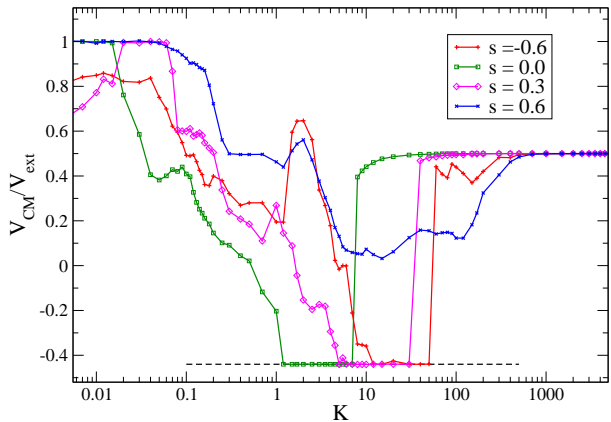}
  \end{center}
\caption{\label{Ak1:fig} (Color online)
  Antikink geometry, for $r=25/36<1$: the average center-mass velocity
  $V_{\rm CM}/V_{\rm ext}$ of the lubricant chain as a function of its
  stiffness $K$.
  Note the negative-velocity plateau, representing a leftward lubricant
  motion.
  The other parameters are $V_{\rm ext}=0.1$, $r'=36/11$, and the model is
  simulated taking $N_+=144$, $N_0=100$ and $N_-=44$, for deformations
  $s=-0.6$, $0.0$, $0.3$, and $0.6$.
  The dashed line marks $w_{\rm plateau}= -0.44$.
}
\end{figure}

When $r<1$, i.e.\ when lubricant particles are fewer than the minima in the
static substrate ($N_0< N_+$), Eq.~\eqref{vplateau} predicts that the
lubricant velocity turns negative, i.e.\ opposite to $V_{\rm ext}$.
This remarkable leftward motion, produced by rightward moving antikinks is
indeed observed even for the deformed RP potential, as shown in the example
of Fig.~\ref{Ak1:fig}.
The resulting ``backward'' plateaus are not so wide as in the case of
kink-assisted forward lubricant motion.
This qualitative finding, quite likely to carry forward to experiment is
due to the dissipation into the substrate represented by the last term in
Eq.~\eqref{model:eq} which tend to favor a positive ``symmetric'' speed
$V_{\rm CM}\simeq V_{\rm ext}/2$, thus actively disturbing the $V_{\rm CM}<0$ quantized
motion.
Like in the $r=\phi$ case, the plateau width and $K$ range depends quite
sensitively on the deformation parameter $s$.
In particular positive $s$ is also especially effective in disrupting the
plateau, and indeed for very strong deformation $s=0.6$ the backward
plateau disappears altogether, the top driving substrate being unable to
grab and drag the antikink pattern in the confined chain.

\subsection{The effect of coverage}

The choices of $r'$ adopted in the calculations of both Figs.~\ref{GM1:fig}
and \ref{Ak1:fig} produce coverage $\Theta=1$, i.e.\ as many solitons (or
antisolitons) as periodic oscillations of the top slider.
As was pointed out \cite{Vanossi07,Castelli08Lyon}, such perfect matching
is the most favorable for kinks (or antikinks) dragging, thus for the
quantized sliding phenomenon.
To investigate how releasing the $\Theta=1$ matching condition affects the
lubricant dynamics, we compare several calculations with the same $r$ but
with different values of the coverage obtained by changing $r'$, i.e.\ the
top substrate lattice spacing $a_-$.

\begin{figure}
 \begin{center}
  \includegraphics[width=80mm,clip=]{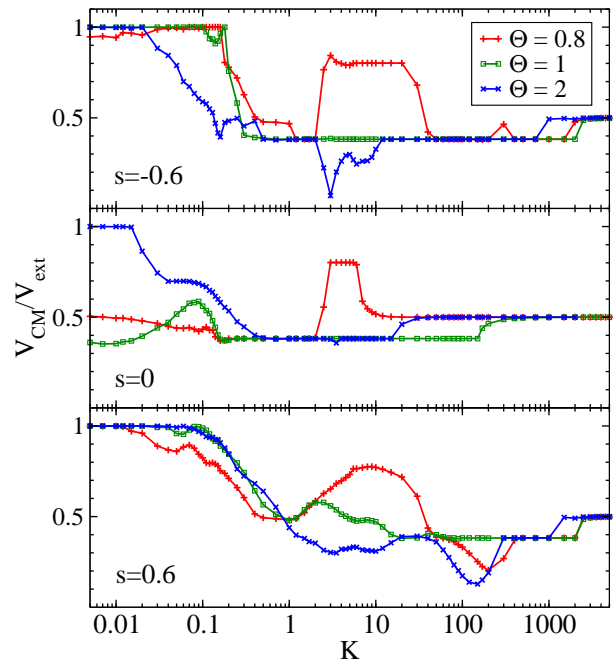}
 \end{center}
\caption{\label{GM6:fig} (Color online)
  Normalized center-mass velocity, $V_{\rm CM}/V_{\rm ext}$ as a function
  of the stiffness $K$, for three values of the deformation parameter.
  The simulations are carried out for $N_0=377$ particles in $N_+=233$
  bottom wells with $N_-=180$, $144$, and $72$, producing coverage $\Theta
  = 0.8$, $1$, and $2$ respectively.
  Note the shrinkage of the plateau for $\Theta\neq 1$, quite substantially
  so for $\Theta =0.8$.
}
\end{figure}

To investigate the coverage dependency, we consider fixed $r = 377/233$ and
the following values of $r'$: $233/180$, $233/144$ and $233/72$
corresponding to kink coverage $\Theta = 0.8$, $\Theta = 1$ and $\Theta =2$
respectively.
The results of these calculations are displayed in Fig.~\ref{GM6:fig} for
three values of the shape parameter.
It can be seen that, independently on the value of the deformation parameter
$s$, the kink coverage largely affects the plateau by reducing its width as
soon as the full matching (unit coverage) is lost.
In general, for the less commensurate the kink coverage $\Theta=0.8$ the
plateau is disrupted quite substantially.
Not surprising, for the commensurate $\Theta=2$ this reduction is less
significant.
Still, for $\Theta=2$ only one kink out of two finds a top corrugation
which drags it along: accordingly, a plateau shrinkage is observed
nonetheless.

We have carried out similar simulations for other commensurability ratios,
e.g.\ the value $r=25/36$ considered in Sect.~\ref{anti:sec}.
The conclusion is that the quantized plateau always shrinks, often to the
point of disappearing, whenever $\Theta \neq 1$.

\section{Conclusions}


In this work, we study the effect of the potential shape on the dynamics of
a sliding 2-substrate FK-type model.
Even though this deformation goes a modest step away from the idealized
world of models in the direction of real friction, it provides some useful
trends and general understanding.
In particular, we establish that in the rigid-lubricant limit (large $K$)
the worst possible pinning scheme for the solitons is that granted by a
sinusoidal corrugation.
Any kind of deformation is beneficial to the quantized state.
This is quite remarkable also in view of the fact that this rigid-lubricant
regime is relevant whenever the lubricant-lubricant in-plane forces
dominate over the substrate corrugation, e.g.\ for noble-gases layers
driven over graphite or even over several metallic surfaces
\cite{Bruch07,Sharma89,Lv11}.
On the opposite, soft-lubricant limit instead a sinusoidal surface
corrugation tends to be optimal, although a pattern based on narrow (but
not too narrow wells may occasionally provide even more favorable
conditions for the quantized lubricant state.

The grabbing of kinks by the more rarefied top slider
is best seen when the kink lattice is fully commensurate to the top.
Whenever this is not the case (kink coverage deviating from unity) the
quantization phenomenon becomes less prominent.
However, this observation is to be integrated by a further point.
As illustrated in the upper and central panels of Fig.~\ref{GM6:fig},
secondary plateaus can arise $V_{\rm CM}$ values different from those predicted
by the quantized formula \eqref{vplateau}.
For example, in the $\Theta =0.8$ simulations producing the secondary
plateau characterized by $V_{\rm CM}/V_{\rm ext} \simeq 0.802$, individual
particles do carry out regular periodic trajectories, like in the standard
quantized state.
These secondary plateaus, observed also for the purely sinusoidal
corrugation \cite{Santoro06,Manini08Erice}, are likely to be due to
resonances very much akin to Shapiro steps
\cite{Floria96,Tekic09,Tekic11,Mali12}, excited by the simultaneous action
of the periodically oscillating force produced by the sliding substrate and
the forward-dragging force produced by the dissipative term in
Eq.~\eqref{model:eq}.
Further investigation of such secondary plateaus can lead to better
insight in their nature, and possibly find realistic configurations where
they could arise, e.g.\ in colloidal sliding
\cite{Bohlein12,Bohlein12PRL,Vanossi12,Vanossi12PNAS}.

\section*{Acknowledgement}

RLW is grateful to the Abdus Salam International Centre for Theoretical
Physics (ICTP) where a part of this work was carried out during her visit
under the associate federation scheme.
The authors thank E. Tosatti and G E Santoro for collaboration,
discussion, and continuing support.
This work was partly sponsored by Sinergia CRSII2$_1$36287/1,
and by advances of ERC Advanced Grant No.\ 320796-MODPHYSFRICT.


\end{document}